\documentclass[12pt]{JHEP3}
\let\ifpdf\relax
\usepackage{amsmath, amscd, amsmath, amsthm, amssymb, xy, xypic}
\usepackage{amssymb}
\usepackage{epsfig}
\usepackage{longtable}
\usepackage{cite}

\newtheorem{theorem}{Theorem}[section]

\newtheorem{condition}[theorem]{Condition}

\numberwithin{equation}{section}

\def \hd #1 {\bfseries #1  \mdseries}
\def \italic #1 {\bfseries \it #1 \rm \mdseries}
\def \ra {\rightarrow}
\def \cen #1 { \begin{center} #1 \end{center}}

\def \mbz {\mathbb Z}

\def \mbc {\mathbb C}

\def \mco  {\mathcal {O}}

\def \Spin {{\rm{Spin}}}
\newcommand{\iu}{{ \sqrt{-1}}}

\ifpdf
\else
\usepackage{colordvi}
\fi

\numberwithin{equation}{section}

\title{A compact manifold with holonomy $\Spin(7)$ from Beauville's Calabi--Yau fourfold}

\author{
Nam-Hoon Lee
\\
Department of Mathematics Education, Hongik University,
42-1, Sangsu-Dong, Mapo-Gu,
Seoul 121-791, South Korea
\\
E-mail: \email{nhlee@hongik.ac.kr},
\\
School of Mathematics, Korea Institute for Advanced Study,
\\
Dongdaemun-gu, Seoul 130-722, South Korea
\\
E-mail: \email{nhlee@kias.re.kr}
}

\abstract{
We give a new example of  a compact manifold with holonomy $\Spin(7)$ from a Beauville's Calabi--Yau fourfold. Its construction is very concrete, starting with  products of elliptic curves with complex multiplications --- so probably  more accessible to physicists.

}

\keywords{ $\Spin(7)$-manifold, F-theory, M-theory, Calabi--Yau fourfold}

\preprint{}

\begin{document}

\section{Introduction}
A $\Spin(7)$-manifold is an eight-dimensional Riemannian manifold with the exceptional holonomy group $\Spin(7)$.
The spin group $\Spin(7)$  is one of special holonomy groups in Berger's classification\cite{Ber}. Riemannian manifolds with special holonomy play an important role in string theory.
Indeed consideration of string theory  compactification on Spin(7) manifolds was proposed by Witten \cite{Wi1, Wi2} and Vafa \cite{Va} two decades ago.
Various aspects of string theory on  $\Spin(7)$-manifolds have been investigated \cite{Bec,GuSp,CuKoLu,AcOsGu,GuSpTo,BeCoGaLiMePh}.
Recently more concrete F-theory approach has been made in \cite{BoGrPu} and \cite{BoGrPaPu}, where the authors  used examples of compact $\Spin(7)$-manifolds that are constructed as quotients of Calabi-Yau fourfolds by Joyce's method \cite{Jo2}.

The construction of  $\Spin(7)$-manifolds
had been an unsolved problem for a long time.
Joyce constructed first compact examples \cite{Jo1}. Later
he gave another method  and constructed  further examples, starting from some complete intersections in weighted projective spaces \cite{Jo2}.
 Following his lead, Clancy, one of his students, systematically investigated hypersurfaces in weighted projective spaces and constructed  more examples \cite{Cl}.
 The Betti numbers of  compact manifolds with holonomy $\Spin(7)$, constructed by them \cite{Jo2, Jo1, Cl}, are
\cen{$0 \leq b^2 \leq 9$,   $0 \leq b^3 \leq 33$, $ 200 \leq b^4 \leq 15118$,}

In Joyce's method, one starts with certain orbifolds, whose resolutions singularities are Calabi-Yau fourfolds.
Since Calabi--Yau fourfolds are projective varieties, it is basically a task in algebraic geometry to find  Calabi--Yau fourfolds suitable for Joyce's method.
One can easily find a huge number of examples of Calabi--Yau fourfolds as complete intersection of toric varieties. The main issues are whether they have suitable singularities and whether they admit antiholomorphic involutions satisfying certain conditions.
Joyce and Clancy considered complete intersections in weighted projective spaces whose antiholomorphic involutions come from those of the ambient weighted projective spaces.

In this note, we apply Joyces's  method to a Calabi--Yau fourfold that is not a  complete intersection in a weighted projective space nor more generally   a  complete intersection in a toric variety.
This Calabi-Yau fourfold was originally constructed by Beaubille \cite{Bea} and is the only one that can be constructed by quotienting an abelian fourfold in a way that Kummer $K3$ surfaces are constructed. It also has rich structure of elliptic fibrations.
As a result, we give a new example of  compact $\Spin(7)$-manifolds. We calculate
the Betti numbers of the example, which are:
\cen{$b^2 = 10$, $b^3 = 30$, $b^4 =52$.}
It is notable that the Betti number $b^4$ of the example  is significantly smaller than those of other examples already constructed. The construction is very concrete, starting with  products of elliptic curves with complex multiplications --- so probably  more accessible to physicists.

\section{Joyce construction from Calabi--Yau 4-orbifolds}
Joyce started with orbifolds with certain conditions \cite {Jo2}. However it is not hard to show that those orbifolds are projective. So let us just start with projective varieties.
Let $Y$ be a 4-dimensional projective varieties satisfying the following conditions.
\begin{condition} \label{cond1}
\begin{enumerate}
\item Each of the singularities of $Y$ is locally isomorphic to  the origin of the quotient $\mbc^4 / \langle \iu* \rangle$, where $\iu*$ acts as the complex multiplication by $\iu$ on $\mbc^4$. Let $p_1, p_2, \cdots, p_k$ ($k \geq 1$) be all the singularities of $Y$.
\item There is  an  antiholomorphic involution $\rho$ on $Y$ whose fixed points  are  $p_1, p_2, \cdots, p_k$.
\item $Y - \{p_1, p_2, \cdots, p_k\}$ is simply-connected.
\item Let $\hat Y \ra Y$ be the blow-up at the singularities of $Y$. Then $\hat Y$ is a Calabi--Yau fourfold, i.e.\ a smooth projective fourfold with trivial canonical class and  $h^{1}(\mco_{\hat Y})=h^{2}(\mco_{\hat Y}) = 0$.
\end{enumerate}
\end{condition}
The final condition may look different from joyce's original one.
However noting the singularities in the condition are crepant, they are actually not different.

Let us consider the quotient $Z = Y/\langle \rho \rangle$. Joyce found a way of  resolving singularities of $Z$ so that the resulted 8-manifolds admits a Riemannian metric whose holonomy group is $\Spin(7)$\cite{Jo2}.
\begin{theorem}[D. Joyce]
There is a simply-connected compact 8-manifold $M$ ( defined in \cite{Jo2}, Definition 5.8) which is a  resolution of singularties $Z = Y/\langle \rho \rangle$ and admits a Riemannian metric whose holonomy group is $\Spin(7)$.
\end{theorem}
The Betti numbers of $M$ can be calculated from topological invariants of $Y$ and $Z$ as follows (Proposition 10, \cite{Cl}):
\begin{align}
b^2(M) &= b^2 (Z), \label{e1} \\
b^3(M) &= \frac{1}{2} b^3 ( Y ), \label{e2}  \\
b^4(M) &= \frac{1}{2} h^{2,2}( Y ) + h^{3,1}( Y) - 2b^2(Z) + \frac{3}{2}k. \label{e3}
\end{align}

\section{Beauville's Calabi--Yau fourfold}

Let
$E=\mbc/(\mbz + \mbz  \sqrt{-1})$ be the elliptic curve with period $ \sqrt{-1}$ and let $Y = E^4/ \langle \iu* \rangle$ be the quotient fourfold of the product manifold $E^4$ by scalar multiplication by $\iu$.
Then $Y$ has finitely many singularities. Let $X \ra Y$ be blow-up at those singularities.
Beauville observed that $X$ is a  simply-connected Calabi--Yau fourfold with Hodge number $h^{1,3} =0$ (\cite{Bea}, page 5).

The fixed points of $E$ by scalar multiplication by $\iu $ are
\cen{$0$, $\alpha = \frac{1 + \iu}{2}$.}
Therefore $Y$ has $2^4 = 16$ singularities and it is easy to see that they are locally isomorphic to  the origin of $\mbc^4 / \langle \iu* \rangle$.

It seems that its Hodge numbers have not been calculated yet. So let us determine all other Hodge numbers of $X$.
Firstly
\cen{$h^{0,0} = 1$, $h^{1,0} = h^{2,0} = h^{3,0} = 0$, $h^{4,0} = 1$.}
The Hodge diamond of $X$ is
\begin{align*}
   & 1\\
0\,\,\,\, &\, \,\,\,\,\, \,\,\, 0\\
0\,\,\,\,\,\,\,\, \,\,\, &h^{3,3} \,\,\,\,\,\,\,\,0\\
0\,\,\,\,\,\,\,\,\,\,h^{2,3} &\, \,\,\, \,\,\,\, h^{3,2}\,\,\,\,\,\,\,\,0\\
1\,\,\,\,\,\,\,\,\,\,h^{1,3} \,\,\,\,\,\,\, &h^{2,2} \,\,\,\,\,\,\,h^{3,1}\,\,\,\,\,\,\,\,\,1\\
0\,\,\,\,\,\,\,\,\,\,h^{1,2} &\, \,\,\, \,\,\,\, h^{2,1}\,\,\,\,\,\,\,\,0\\
0\,\,\,\,\,\,\,\, \,\,\, &h^{1,1} \,\,\,\,\,\,\,\,0\\
0\,\,\,\, &\, \,\,\,\,\, \,\,\, 0\\
   & 1\\
\end{align*}

with
\cen{$h^{1,1} = h^{3,3}$, $h^{3,1} = h^{1,3}$, $h^{1,2}=h^{2,1} = h^{3,2}=h^{2,3}$.}
Let $\chi_q = \sum_{p=1}^4 (-1)^p h^{p,q}$, then
by the well-known Riemann--Roch theorem, we have
\begin{align*}
\chi_0 &= \frac{1}{720}(-c_4 + c_1 c_3 + 3 c_2^2 + 4 c_1^2 c_2 - c_1^4)\\
\chi_1 &= \frac{1}{180} ( - 31 c_4 - 14 c_1 c_3 + 3 c_2^2 + 4 c_1^2 c_2 - c_1^4)\\
\chi_2 &= \frac{1}{120} ( 79 c_4 - 19 c_1 c_3 + 3 c_2^2 + 4 c_1^2 c_2 - c_1),
\end{align*}
where $c_i$ is the $i$th Chern class of $X$. Since $X$ is a Calabi--Yau fourfold, $c_1 = 0$ and $\chi_0 = 2$.  With this, we have following relations in Hodge numbers:
$$h^{2,2} = 2(22 + 2h^{1,1} + 2h^{3,1} - h^{2,1}).$$
So remaining independent Hodge numbers are $h^{1,1}$ and $h^{1,2}$.
The topological Euler number of $X$ has the relation:
\begin{align}
e(X) = 6(8 + h^{1,1} + h^{3,1} - h^{1,2}). \label{eqn1}
\end{align}
Let $\widetilde{E^3} \ra E^3$ be the blow-up at $\{0, \alpha \}^4$ and $F_{ijkl}$ be the exceptional divisors over $(e_i, e_j, e_k, e_l)$ for $i, j = 0, 1$, where $e_0 = 0$ and $e_1 = \alpha$.
Then there is a quadruple covering map $X \ra \widetilde{E^4}$, branched along $F_{ijkl}$'s and we have the following commutative diagram:

$$
\xymatrix{
X \ar[rr] \ar[dd]   & &   \widetilde{E^4}  \ar[dd]\\
&&\\
Y    \ar[rr]   && E^4
}
$$

The topological Euler numbers are
        $$e(E^4)=0,$$
$$e \left (\widetilde{E^4} \right )  = e(E^4) + \sum_{i,j,k,l} \left (e(F_{ijkl})-1 \right) = 2^4 \cdot 3,$$
$$4 e  (X )  - 3 \cdot \sum_{i,j,k,l}e(F_{ijkl}) =  e \left (\widetilde{E^4} \right ).$$
So we have
$$ e  (X ) =60.$$
Hence Equation \ref{eqn1} becomes
$$60 = 6(8 + h^{1,1} + 0 - h^{1,2}).$$
Therefore
\begin{align}
h^{1,1} - h^{1,2} = 2. \label{starstar}
\end{align}

On the other hand, note (\cite{BiLa}, page 21)
\begin{align}
h^{1,1}(E^4) = 4^2= 16. \label{star}
\end{align}
Now let us find  generators of $H^{1,1}(E^4)$.
Let $\pi_i: E^4 \ra E$ be the $i$-the projection.
Consider following $16$ divisors (denoted by $b_i$'s)  of $E^4$:
\begin{itemize}
\item  $\beta_i = \ker \pi_i$ ($=b_i$) for $i=1,2,3,4$.
\item  $\gamma_{ij} = \ker (\pi_i + \pi_j)$ for $1 \leq i < j \leq 4$.
Let
\cen{$b_5=\gamma_{1\,2}$, $b_6=\gamma_{1\,3}$, $b_7=\gamma_{1\,4}$, $b_8=\gamma_{2\,3}$, $b_9=\gamma_{2\,4}$, $b_{10}=\gamma_{3\,4}$.}
\item  $\delta_{ij} = \ker (\pi_i + \iu \,\pi_j)$ for $1 \leq i < j \le 4$.
Let
\cen{$b_{11}=\delta_{1\,2}$, $b_{12} =\delta_{1\,3}$, $b_{13} =\delta_{1\,4}$, $b_{14}=\delta_{2\,3}$, $b_{15}=\delta_{2\,4}$, $b_{16} =\delta_{3\,4}$.}
\end{itemize}
These $b_i$'s can be regarded as elements of $H^{1,1}(E^4 )$.

Now consider $16$ elements (denoted by $c_j$'s) of $H^{3,3}(E^3)$:
\begin{itemize}
\item $\beta_i \cdot \beta_j  \cdot \beta_k$ for $ 1\leq i<j<k \leq 4$.
Let
\cen{$c_1=\beta_1 \cdot \beta_2\cdot \beta_3$, $c_2=\beta_1 \cdot \beta_2\cdot \beta_4$, $c_3=\beta_1 \cdot \beta_3\cdot \beta_4$, $c_4=\beta_2 \cdot \beta_3\cdot \beta_4$.}
\item $\gamma_{ij} \cdot \beta_k \cdot \beta_l$ for $ 1 \leq i<j \leq 4, 1 \leq k<l \leq 4$ and $\{i, j\} \neq \{ k, l \}$.
    Let
\cen{$c_5=\gamma_{1\,2} \cdot \beta_3\cdot \beta_4$, $c_6=\gamma_{1\,3} \cdot \beta_2\cdot \beta_4$, $c_7=\gamma_{1\,4} \cdot \beta_2\cdot \beta_3$, $c_8=\gamma_{2\,3} \cdot \beta_1\cdot \beta_4$, $c_{9}=\gamma_{2\,4} \cdot \beta_1\cdot \beta_3$, $c_{10}=\gamma_{3\,4} \cdot \beta_1\cdot \beta_2$.}
\item $\delta_{ij} \cdot \beta_k \cdot \beta_l$ for $ 1 \leq i<j \leq 4, 1 \leq k<l \leq 4$ and $\{i, j\} \neq \{ k, l \}$.
    Let
\cen{$c_{11}=\delta_{1\,2} \cdot \beta_3\cdot \beta_4$, $c_{12}=\delta_{1\,3} \cdot \beta_2\cdot \beta_4$, $c_{13}=\delta_{1\,4} \cdot \beta_2\cdot \beta_3$, $c_{14}=\delta_{2\,3} \cdot \beta_1\cdot \beta_4$, $c_{15}=\delta_{2\,4} \cdot \beta_1\cdot \beta_3$, $c_{16}=\delta_{3\,4} \cdot \beta_1\cdot \beta_2$,}
\end{itemize}
where `$\cdot$' is the cup product.
Let $N$ be the $16 \times 16$ intersection matrix of $b_i$'s and $c_j$'s (i.e.\ $N_{ij} = b_i \cdot c_j$). Then $N$ is as follows:

 $$N = {\left(
\begin{array}{rrrrrrrrrrrrrrrr}
 0 & \,\,0 & \,\,0 & \,\,1 & \,\,1 & \,\,1 & \,\,1 & \,\,0 & \,\,0 & \,\,0 & \,\,1 & \,\,1 & \,\,1 & \,\,0 & \,\,0 & \,\,0 \\
 0 & 0 & 1 & 0 & 1 & 0 & 0 & 1 & 1 & 0 & 1 & 0 & 0 & 1 & 1 & 0 \\
 0 & 1 & 0 & 0 & 0 & 1 & 0 & 1 & 0 & 1 & 0 & 1 & 0 & 1 & 0 & 1 \\
 1 & 0 & 0 & 0 & 0 & 0 & 1 & 0 & 1 & 1 & 0 & 0 & 1 & 0 & 1 & 1 \\
 0 & 0 & 1 & 1 & 0 & 1 & 1 & 1 & 1 & 0 & 2 & 1 & 1 & 1 & 1 & 0 \\
 0 & 1 & 0 & 1 & 1 & 0 & 1 & 1 & 0 & 1 & 1 & 2 & 1 & 1 & 0 & 1 \\
 1 & 0 & 0 & 1 & 1 & 1 & 0 & 0 & 1 & 1 & 1 & 1 & 2 & 0 & 1 & 1 \\
 0 & 1 & 1 & 0 & 1 & 1 & 0 & 0 & 1 & 1 & 1 & 1 & 0 & 2 & 1 & 1 \\
 1 & 0 & 1 & 0 & 1 & 0 & 1 & 1 & 0 & 1 & 1 & 0 & 1 & 1 & 2 & 1 \\
 1 & 1 & 0 & 0 & 0 & 1 & 1 & 1 & 1 & 0 & 0 & 1 & 1 & 1 & 1 & 2 \\
 0 & 0 & 1 & 1 & 2 & 1 & 1 & 1 & 1 & 0 & 0 & 1 & 1 & 1 & 1 & 0 \\
 0 & 1 & 0 & 1 & 1 & 2 & 1 & 1 & 0 & 1 & 1 & 0 & 1 & 1 & 0 & 1 \\
 1 & 0 & 0 & 1 & 1 & 1 & 2 & 0 & 1 & 1 & 1 & 1 & 0 & 0 & 1 & 1 \\
 0 & 1 & 1 & 0 & 1 & 1 & 0 & 2 & 1 & 1 & 1 & 1 & 0 & 0 & 1 & 1 \\
 1 & 0 & 1 & 0 & 1 & 0 & 1 & 1 & 2 & 1 & 1 & 0 & 1 & 1 & 0 & 1 \\
 1 & 1 & 0 & 0 & 0 & 1 & 1 & 1 & 1 & 2 & 0 & 1 & 1 & 1 & 1 & 0
\end{array}
\right) } $$

The rank of the matrix $N$ is $16$, which means that $b_i$'s and $c_j$'s are linearly independent respectively. Since $h^{1,1}(E^4) = h^{3,3}(E^4) =16$ (Equation \ref{star}),  they form  bases of $H^{1,1}(E^4)$ and $H^{3,3}(E^4)$ respectively.
 Moreover $b_i$' and $c_j$'s are all invariant under the  scalar multiplication by $\iu$.
 Since elements of $H^{1,1}(Y)$ come from cycles in $H^{1,1}(E^4)$ that are invariant under the scalar multiplication by $\sqrt{-1}$, we conclude that $h^{1,1}(Y) = 16$
and accordingly
$$h^{1,1}= h^{1,1}(Y)+ 16 = 32.$$
By Equation \ref{starstar}, we have  $h^{1,2}=30$ and
$$h^{2,2} = 2(22 + 64  - 30) = 112.$$

\section{An example of compact manifold with holonomy $\Spin(7)$ }
Now let us find a suitable antiholomorphic involution on $Y$.
Let $c:E^4 \ra E^4$ be the standard complex conjugation, i.e.
$$c: \left(
                           \begin{array}{c}
                             z_1 \\
                             z_2 \\
                             z_3 \\
                             z_4 \\
                           \end{array}
                         \right)
 \mapsto
 \left(
                           \begin{array}{c}
                             \bar z_1 \\
                             \bar z_2 \\
                             \bar z_3 \\
                             \bar z_4 \\
                           \end{array}
                         \right)$$
and $A$ be a $4 \times 4$  matrix which takes its entries from $\mbz[\sqrt{-1}]$. Note that $A$ induces a holomorphic map $\widehat A : E^4 \ra E^4$. Let $\psi_A = \widehat A \circ c$, then it is an
antiholomorphic map and $\psi_A$ also induces an antiholomorphic map $\phi_A : Y \ra Y$.  Note
$$\psi_A^2 = \widehat A \circ c \circ \widehat A \circ c = \widehat A \circ \widehat {\overline A}  = \widehat {A \overline A},$$
where $\overline A$ is the $4 \times 4$ matrix whose entries are complex conjugations of those of $A$.
So $\phi_A$ is an involution if and only if $A \overline A = I$, $-I$, $\sqrt{-1} I$ or $-\sqrt{-1} I$, where $I$ is the $4 \times 4$ identity matrix.
Let
$$A=\left(
          \begin{array}{cccc}
            -1 & 1+\iu & 0 & 0 \\
            -1-\iu & \iu & 0 & 0 \\
            0 & 0 & -1 & 1+\iu \\
            0 & 0 & -1-\iu & \iu \\
          \end{array}
        \right)
$$
Then $A \overline A = -I$. So $\phi_A$ is an antiholomorphic involution of $Y$.

By direct calculation, one can show that fixed points of $\phi_A$ are exactly the singularities of $Y$.
So we can apply Joyce's method to $Y$ with the antiholomorphic involution $\phi_A$ to get a compact $\Spin(7)$-manifold $A$.
In order to determine the Betti numbers of $M$, we need to calculate the Betti number $b^2(Z)$.

By calculating the cup product numbers $c_i \cdot \psi_A(b_j)$'s,
one can  find the $16 \times 16$ matrix that represents how $\psi_A$ works on $H^{1,1}(E^4)$ with respect to the basis $\{b_i's \}$ and it is:

$$
\left(
\begin{array}{rrrrrrrrrrrrrrrr}
 -1 & -2 & 0 & 0 & \,\,\,\,1 & 0 & \,\,\,\,0 & \,\,\,\,0 & \,\,\,\,0 & \,\,\,\,0 & -1 & 0 & 0 & 0 & 0 & 0 \\
 -2 & -1 & 0 & 0 & 1 & 0 & 0 & 0 & 0 & 0 & -1 & 0 & 0 & 0 & 0 & 0 \\
 0 & 0 & -1 & -2 & 0 & 0 & 0 & 0 & 0 & 1 & 0 & 0 & 0 & 0 & 0 & -1 \\
 0 & 0 & -2 & -1 & 0 & 0 & 0 & 0 & 0 & 1 & 0 & 0 & 0 & 0 & 0 & -1 \\
 -6 & -6 & 0 & 0 & 4 & 0 & 0 & 0 & 0 & 0 & -3 & 0 & 0 & 0 & 0 & 0 \\
 0 & -2 & -2 & 0 & 1 & -1 & 1 & 1 & -2 & 1 & -1 & 0 & -1 & 1 & 0 & -1 \\
 0 & -4 & -4 & 0 & 1 & -1 & 0 & 2 & -1 & 1 & -1 & 1 & -1 & 0 & 1 & -1 \\
 -2 & 0 & 0 & -2 & 1 & -1 & 2 & 0 & -1 & 1 & -1 & -1 & 0 & 1 & -1 & -1 \\
 0 & -2 & -2 & 0 & 1 & -2 & 1 & 1 & -1 & 1 & -1 & 0 & -1 & 1 & 0 & -1 \\
 0 & 0 & -6 & -6 & 0 & 0 & 0 & 0 & 0 & 4 & 0 & 0 & 0 & 0 & 0 & -3 \\
 -2 & -2 & 0 & 0 & 1 & 0 & 0 & 0 & 0 & 0 & 0 & 0 & 0 & 0 & 0 & 0 \\
 0 & -4 & -2 & -2 & 1 & 0 & -1 & 1 & 0 & 1 & -1 & 1 & -1 & -1 & 2 & -1 \\
 -2 & -2 & -2 & -2 & 1 & 1 & -1 & 0 & 1 & 1 & -1 & 1 & 0 & -2 & 1 & -1 \\
 0 & -2 & -2 & 0 & 1 & -1 & 0 & 1 & -1 & 1 & -1 & 1 & -2 & 0 & 1 & -1 \\
 -2 & -2 & -4 & 0 & 1 & 0 & -1 & 1 & 0 & 1 & -1 & 2 & -1 & -1 & 1 & -1 \\
 0 & 0 & -2 & -2 & 0 & 0 & 0 & 0 & 0 & 1 & 0 & 0 & 0 & 0 & 0 & 0
\end{array}
\right)
$$
 The dimension of the eigenspace of the matrix to the eigenvalue one can be shown to be $10$.
 So $b^2(Z)=10$ and
By Equation \ref{e1}, \ref{e2}, \ref{e3}, we have
\begin{align*}
b^2(M) &= b^2 (Z) = 10, \\
b^3(M) &= \frac{1}{2} b^3 ( Y )=  h^{1,2}(Y) = 30,  \\
b^4(M) &= \frac{1}{2} h^{2,2}( Y ) + h^{3,1}( Y) - 2b^2(Z) + \frac{3}{2}k\\
       &= \frac{1}{2} \cdot 98  + 0 - 2 \cdot 10 + \frac{3}{2} \cdot 16 \\
       &= 52.
\end{align*}

In summary, \cen{$b^2(M) = 10$, $b^3(M) = 30$, $b^4(M) =52$.}
The Betti numbers of  compact manifolds with holonomy $\Spin(7)$, constructed so far  \cite{Jo1, Jo2, Cl}, are
\cen{$0 \leq b^2 \leq 9$,   $0 \leq b^3 \leq 33$, $ 200 \leq b^4 \leq 15118$,}
It is notable that the Betti number $b^4(M)$  is significantly smaller than those of other examples already constructed.

Let us consider more general antihomorphic involutions of $Y$.
Let $\xi = (\xi_1, \xi_2, \xi_3, \xi_4) \in E^4$ with  $\xi_i = 0$ or  $\alpha$.
Then translation $t_\xi$ of $E^4$ by $\xi$ is lifted to an automorphism $\Gamma_\xi$ of $Y$.
In general case, an antiholomorphic automorphism of $Y$ has the form
$$\phi_{B, \xi} := \Gamma_\xi \circ \phi_B,$$
where $B$ is a $4 \times 4$ matrix with entries in $\mbz[\iu]$.

If $\phi_{B, \xi}$ is an involution and fixes some of singularities of $Y$.
Now let $\hat Y \ra Y$ be the blow-ups of $Y$ at its singularities that are not fixed by $\phi_{B, \xi}$. Then $\phi_{B, \xi}$ induces an antiholomorphic involution on $\hat Y$.
It is easy to check that $\hat Y$ with this involution satisfies Condition \ref{cond1}.
The author tested various $B$'s, $\xi$'s  basically by a computer. However he only found only the examples which give the same Betti numbers presented in the previous example.

\acknowledgments{This work was supported by Basic Science Research Program through the National Research Foundation of Korea(NRF) funded by the Ministry of  Science, ICT \& Future Planning (2012R1A1A1039764) and 2012 Hongik University Research Fund.}


\end{document}